\documentclass[fleqn,usenatbib]{mnras}

\usepackage{newtxtext,newtxmath}

\usepackage[T1]{fontenc}

\DeclareRobustCommand{\VAN}[3]{#2}
\let\VANthebibliography\thebibliography
\def\thebibliography{\DeclareRobustCommand{\VAN}[3]{##3}\VANthebibliography}

\usepackage{graphicx}	
\usepackage{amsmath}	
\usepackage{verbatim}
\usepackage{gensymb}    

\title[Planet theft and capture in star-forming regions]{The Great Planetary Heist: Theft and capture in star-forming regions}

\author[E. C. Daffern-Powell, R. J. Parker \& S. P. Quanz]{
Emma C. Daffern-Powell,$^{1}$\thanks{E-mail: ecdaffern1@sheffield.ac.uk}
Richard J. Parker$^{1}$\thanks{E-mail: R.Parker@sheffield.ac.uk}\thanks{Royal Society Dorothy Hodgkin Fellow}
and Sascha P. Quanz$^{2}$
\\
$^{1}$Department of Physics and Astronomy, The University of Sheffield, Hicks Building, Hounsfield Road, Sheffield S3 7RH, UK\\
$^{2}$Institute for Particle Physics and Astrophysics, ETH Z{\"u}rich, Wolfgang-Pauli-Strasse 27, CH-8093 Z{\"u}rich, Switzerland
}

\pubyear{2021}

\begin{document}
\label{firstpage}
\pagerange{\pageref{firstpage}--\pageref{lastpage}}
\maketitle

\begin{abstract}
Gravitational interactions in star-forming regions are capable of disrupting and destroying planetary systems, as well as creating new ones.
In particular, a planet can be stolen, where it is directly exchanged between passing stars during an interaction; 
or captured, where a planet is first ejected from its birth system and is free-floating for a period of time, before being captured by a passing star.
We perform sets of direct $N$-body simulations of young, substructured star-forming regions, and follow their evolution for 10 Myr in order to determine how many planets are stolen and captured, and their respective orbital properties.
We show that in high density star-forming regions, stolen and captured planets have distinct properties. The semimajor axis distribution of captured planets is significantly skewed to wider orbits compared to the semimajor axis distribution of stolen planets and planets that are still orbiting their parent star (preserved planets). However, the eccentricity and inclination distributions of captured and stolen planets are similar, but in turn very different to the inclination and eccentricity distributions of preserved planets. 
In low-density star-forming regions these differences are not as distinct but could still, in principle, be used to determine whether observed exoplanets have likely formed in situ or have been stolen or captured. We find that the initial degree of spatial and kinematic substructure in a star-forming region is as important a factor as the stellar density in determining whether a planetary system will be altered, disrupted, captured or stolen. 

\end{abstract}




\section{Introduction} 
Numerous exoplanets have now been observed on orbits that cannot be fully explained by the current theories of star and planet formation, whether that be core accretion \citep{1996PollackEtAl} or the fragmentation of stellar disks \citep{1997Boss, 2002MayerEtAl}. 
For example, dozens of possibly planetary mass companions have been observed with unexpectedly high eccentricities and semimajor axes, with some orbiting as far as $\gtrapprox$2500 AU from their host star \citep[e.g.][]{2011LuhmanEtAl, 2016DeaconEtAl}.
Furthermore, although the abundance of free-floating planets is uncertain \citep{2011SumiEtAl, 2012QuanzEtAl, 2017ClantonGaudi, 2017MrozEtAl, 2019OGLECollab}, several likely candidates have been discovered \citep[e.g.][]{2000Zapatero, 2013DupuyKraus, 2019OGLECollab, 2020MrozEtAl}.

Meanwhile, in our own Solar System, should the proposed Planet 9 exist \citep{2016SheppardTrujillo, 2016BatyginBrown, 2019BatyginEtAl, 2020ClementKaib, 2020DowneyMorbidelli}, its orbit is likely to be very wide ($a \sim 400-800$ AU), eccentric ($e \sim 0.2-0.5$), and inclined ($i \sim 15-25 ^{\circ}$) to the plane of the other planets \citep{2019BatyginEtAl, 2020FiengaEtAl}.
(However, see also \citet{2017ShankmanEtAl} and \citet{2021NapierEtAl} for a more pessimistic view regarding the likelihood of Planet 9's existence.)

Planetary orbits that cannot be explained by core accretion or gravitational fragmentation, as well as free-floating planets, can instead be created by encounters with other bodies \citep[e.g.][]{1998LaughlinAdams, 2002HurleyShara, 2001SmithBonnell, 2006AdamsEtAl, 2012ParkerQuanz, 2015ZhengEtAl, 2016KouwenhovenEtAl, 2019DottiEtAl, 2019LiEtAl, 2020LiEtAl, 2020WangEtAl_architectures}. 
These dynamical interactions with external bodies have the potential to perturb, destabilise, and destroy planetary orbits -- with the main outcome being that the planet's orbital parameters are changed.

Dynamical interactions readily occur in high density environments, where close encounters between systems can be common.
This is particularly the case for the substructured and filamentary regions in which most stars and their planets form \citep{2003LadaLada, 2004CartwrightWhitworth, 2014AndreEtAl, 2020DaffernPowellParker}.
These star-forming regions tend to be relatively dense compared to the galactic field (which has a stellar density of 0.1 pc$^{-3}$, \citealp{2003KorchaginEtAl}). 
For example, Taurus has a stellar density of $\approx5$ stars pc$^{-3}$, and the most massive star-forming regions can have $\gtrapprox1000$ stars pc$^{-3}$, with the Orion Nebula Cluster having a central density of $\approx5000$ stars pc$^{-3}$ \citep{2012KingEtAl}.

It is unclear what fraction of planet-hosting stars form in such dense environments, and whether our Sun formed in a region of high or low-density.
However, there is evidence from isotopic ratios that the Solar System may have formed in a relatively dense cluster of $\sim100$ to $\sim1000$ stars \citep{2010Adams, 2016ParkerDale, 2016LichtenbergEtAl, 2017NicholsonParker, 2019Zwart}.

The substructured nature of star-forming regions can also further increase the frequency of dynamical interactions, as areas of substructure have higher local densities than the global density of the region as a whole \citep{2004CartwrightWhitworth, 2009SanchezAlfaro, 2010AndreEtAl, 2012ParkerMeyer, 2014AndreEtAl, 2014KuhnEtAl, 2015JaehnigEtAl, 2019ArzoumanianEtAl, 2020BalloneEtAl}.

The dynamical interactions that planets are more likely to experience in star-forming regions can be grouped by mechanism.
These mechanisms include
the \emph{disruption} of a planet's orbit, where one or more of its semimajor axis, eccentricity, or inclination is altered \citep{2002HurleyShara, 2009SpurzemEtAl, 2012ParkerQuanz};
the \emph{ejection} of a planet from its system to become free-floating, where it is no longer gravitationally bound to a star \citep{2019FujiiHori, 2019VanElterenEtal, 2019CaiEtAl};
the \emph{capture} of a free-floating planet by either a new star or its original star \citep{2012PeretsKouwenhoven, 2015WangEtAl, 2017ParkerEtAl};
and the \emph{theft} of a planet, where it is directly exchanged between stars as they pass each other\footnote{As an aside, a vast amount of literature exists on the exchange of disc material between stars \citep[e.g.][]{Clarke1993,Kenyon2004,Jilkova2015,Li2020,Pfalzner2021}, which is the same physical mechanism that accounts for the theft of a fully-formed planet from a passing star.}, without being free-floating for a significant period of time\footnote{We take a significant period of time to be $10^4$ years, as described in \S\ref{sec:orbit_types}.} \citep{2016LiAdams, 2016MustillEtAl, 2020WangEtAl_swaps}.

The result of planet theft and the capture of a free-floating planet is often the same -- a planet orbiting a new star.
However, the two mechanisms are themselves distinct and the evolution of the planets that undergo them may also differ significantly.
For example, a free-floating planet may experience encounters that change its velocity before its subsequent capture \citep{2015WangEtAl}, something which cannot have happened to a stolen planet.
It is therefore unclear whether stolen and captured planets should be expected to share similar orbital properties, or whether they would be distinguishable observationally.

The current state-of-the-art in modelling planetary systems in $N$-body simulations of star-forming regions is to effectively run simulations of planetary systems within the global star-forming region simulations; interaction histories between stars are tracked and then used to determine the amount of disruption experienced by multi-planet systems using separate software \citep[e.g.][]{2019CaiEtAl,2019DottiEtAl,Stock20}. The advantage of this approach is a full planetary system can be modelled (instead of just one or two planets) and that the long-term dynamical evolution of the planets can be accurately determined. The disadvantage is that any planet(s) that are ejected from their system cannot re-enter the global simulation of the star-forming region as a free-floating planet.

  Furthermore, to date, the simulations of the star-forming regions with multi-planet systems assume smooth, relaxed initial conditions. However, planets form almost immediately after the onset of star formation \citep{Brogan15,Andrews18,Alves20,SeguraCox20}, when star-forming regions exhibit a large degree of spatial and kinematic substructure. If the star-forming regions have low densities initially, this substructure can last for many crossing times (i.e.,\,\,several Myr) and so there is merit in examining the effects of substructured star-forming regions on young planetary systems \citep{2012ParkerQuanz,2013CraigKrumholz}. Indeed, \citet{Parker21a} demonstrated that external photoevaporation of protoplanetary discs is slightly hindered in more substructured regions, as the massive stars are on average further away from the majority of disc-bearing stars, even though the median stellar density is the same as in regions with less substructure.

In this paper, we perform direct $N$-body simulations of dense, substructured star-forming regions, where half of the stars have Jupiter-mass planets placed at 30 or 50 AU. We adopt these orbital distances to facilitate a direct comparison with our previous work \citep{2012ParkerQuanz} and because a significant number of exoplanets are observed at these separations. 
The evolution of all of the planets in each simulation are followed for 10 Myr.
We test whether the orbital properties and  abundances of captured and stolen planets differ significantly from that of each other, as well as that of planets that are still orbiting their parent star.
We then compare these results to directly imaged exoplanets and the hypothetical Planet 9.

In Section~\ref{sec:Methods} we outline our methods of simulation and analysis as well as our adopted initial conditions, in Section~\ref{sec:ResultsDiscussion} we presents our results and discuss their implications, and we conclude in Section~\ref{sec:Conlcusion}.

\section{Methods}
\label{sec:Methods}
In this section we describe the set-up of our simulated star-forming regions and how the results were then analysed.

\subsection{Simulation Set-Up} 
Direct $N$-body simulations were run using the \texttt{kira} $N$-body integrator \citep{1999PortegiesZwart, 2001PortegiesZwart}. These were evolved for 10 Myr, with snapshots of data taken every 0.01 Myr for analysis.
We use several sets of initial conditions, all of which contain 1000 stars, $N_{\star}$, and 500 planets, $N_{\rm p}$, which are randomly assigned to the stars.
The planets are all Jupter-mass.
The planet mass would not be expected to significantly affect our results, especially with respect to the ejection of a planet from its system, as the interaction cross-section is primarily dependant on the stellar mass \citep{2004FregeauEtAl, 2006FregeauEtAl, 2013ParkerReggiani}.

For our main collection of simulations, the stars are placed within a 1 pc region and distributed according to the box-fractal method. The box-fractal method was introduced by \cite{2004GoodwinWhitworth}, and is a commonly used method to set up spatial and velocity structure in $N$-body simulations \citep[e.g.][]{2004GoodwinWhitworth, 2010AllisonEtAl, 2012ParkerQuanz, 2014ParkerEtAl, 2020DaffernPowellParker}.

The spatial substructure is set up as follows:
\begin{enumerate}
    \item A cube with sides of length $N_{\rm div}=2$ is defined. It is within this cube that the star-forming region is generated. The first `parent' particle is placed at the centre of the cube.
    \item The cube is divided into sub-cubes, each with length 1. So, in this case, there are $N_{\rm div}^3=8$ sub-cubes. A `child' particle is placed at the centre of each sub-cube.
    \item The probability that a child particle now becomes a parent itself is $N_{\rm div}^{D-3}$, where $D$ is the fractal dimension. We adopt $D = 1.6$.
    \item Child particles that do not become parents themselves are removed as well as all of their previous generations of parent particles. Children that do become parents have a small amount of noise added to their distribution to prevent a gridded appearance.
    \item Each new generation of parent particles is treated in the same way as the initial parent particle, and their sub-cubes are treated as the initial cube. In this way, each new parents' sub-cube is divided into $N_{\rm div}^3=8$ as the process is repeated until there is a generation created that has significantly more particles than is needed.
    \item Any remaining parent particles are removed so that only the last generation of particles is left.
    \item The region is pruned so that the particles sit within a spherical boundary, with the chosen diameter of 1 or 5 pc, rather than a cube.
    \item If there are more particles than the chosen number of stars, in this case $N_{\star} = 1000$, particles are removed at random until $N_{\star}$ is reached. Removing stars at random maintains the chosen fractal dimension as closely as possible.
\end{enumerate}

The mean number of children that become parents is $N_{\rm div}^{D}$. This means that, when $N_{\rm div} = 2$, fractal dimensions of $D =$ 1.6, 2.0, 2.6, and 3.0 correspond to the mean number of new parents at each stage being close to an integer. This is preferred because it produces the chosen fractal dimension more accurately. Lower fractal dimensions lead to fewer children becoming parents, and therefore more substructure. We adopt a fractal dimension $D = 1.6$ in most of the simulations which corresponds to the maximum amount of substructure possible. In two simulations, we keep the stellar density constant, but change the initial degree of substructure so that $D = 2.0$ (a moderate amount of substructure) or $D=3.0$ (no substructure). To ensure the densities are commensurate with the $D=1.6$ simulations, the radii for the moderately substructured, and zero substructured simulations are 0.5\,pc and 0.25\,pc, respectively (the $D=1.6$ simulations have radii of 1\,pc). 

To set up the velocity substructure:
\begin{enumerate}
    \item The first parent particle has its velocity drawn from a Gaussian with mean of zero.
    \item Every particle after that has the velocity of its parent plus an additional random velocity component. This additional component is drawn from the same Gaussian and multiplied by $(\frac{1}{N_{\rm div}})^{g}$, where $g$ is the number of the generation that the particle was produced in. This results in the additional components being smaller on average with each successive generation of particles.
    \item The velocities are then scaled so that the region has the required virial ratio.
\end{enumerate}

Setting up the velocity substructure in this way ensures that stars that are closer together have more similar velocities than those that are further apart, as is expected from observations, e.g. the Larson relation \citep{1981Larson}.

Our default fractal dimension of 1.6 corresponds to a high amount of substructure, and with an initial radius of 1 pc this leads to an initial median density of order $10^4$ M$_{\odot}$pc$^{-3}$.
We also run an additional set of simulations with an initial radius of 5 pc, which corresponds to an initial median density of order $100$ M$_{\odot}$pc$^{-3}$.

Our combinations of initial conditions are summarised in Table~\ref{tab:initialConditions}, which shows the initial planetary semimajor axes, $a_p$, and virial ratio, $\alpha = T / |\Omega|$, where $T$ is the total kinetic energy and $\Omega$ is the total potential energy of the region.
In this way we investigate the effects of different global motions of the star-forming region with the virial ratio, and different planetary orbits with the semimajor axis.

We use two initial virial ratios of $\alpha=0.3$ and $\alpha=1.5$.
A virial ratio of $\alpha=0.3$ is used to model the subvirial collapse of a region to form a bound cluster \citep{2014ParkerEtAl, 2015FosterEtAl}.
A virial ratio of $\alpha=1.5$ models the supervirial expansion of a region -- corresponding to a region that either formed unbound, or has become unbound e.g. by tidal forces or gas expulsion \citep{1978Tutukov, 1997Goodwin, 2007BaumgardtKroupa}. 

These two virial ratios are combined with three types of initial planetary orbits: orbiting a star with a semimajor axis of $a_p = 30$ AU, orbiting a star with $a_p = 50$ AU, or free-floating. In the context of the Solar System these two semi-major axes correspond to the semimajor axis of Neptune and the outer edge of the Kuiper belt. The star-forming regions that contain free-floating planets are only simulated with a virial ratio of $\alpha=0.3$ \citep[for simulations with free-floating planets and supervirial initial conditions see][]{2012PeretsKouwenhoven, 2017ParkerEtAl}.

All planets that are bound to a star initially have zero eccentricity, $e = 0$, and are orientated randomly with respect to the coordinate system of the simulation i.e. they have random inclinations and random argument of latitude.

\begin{table}
	\centering
	\caption{Summary of the initial conditions. Column 1 indicates whether the planets are initially bound to a host star or free-floating. Column 2 gives the initial radius of the star-forming region.  Column 3 gives the fractal dimension, $D$ of the region.  Column 4 gives the median initial stellar density, $\tilde{\rho}$, column 5 gives the initial virial ratio, $\alpha$, of the star-forming region, and finally, column 6 gives the semimajor axis, $a_p$, of each planetary system.}
	\label{tab:initialConditions}
	\begin{tabular}{l c c c c c}
		\hline
		Planetary Orbit Type  &  $r$  &  $D$ & $\tilde{\rho}$ & $\alpha$  &  $a_p$  \\
		\hline 
		Bound                     &  1\,pc     & 1.6   & $10^4$\,M$_\odot$\,pc$^{-3}$    &  0.3          & 30\,au            \\
                &  0.5\,pc     & 2.0    & $10^4$\,M$_\odot$\,pc$^{-3}$   &  0.3          & 30\,au            \\
                &  0.25\,pc     & 3.0    & $10^4$\,M$_\odot$\,pc$^{-3}$   &  0.3          & 30\,au            \\
	                        &  1\,pc   & 1.6     & $10^4$M\,$_\odot$\,pc$^{-3}$     &  0.3          & 50\,au          \\
		                    &  1\,pc      & 1.6   & $10^4$\,M$_\odot$\,pc$^{-3}$    & 1.5           & 30\,au           \\
		                     &  1\,pc     & 1.6    & $10^4$\,M$_\odot$\,pc$^{-3}$    &  1.5          & 50\,au           \\
							     & 5\,pc     & 1.6    & $10^2$\,M$_\odot$\,pc$^{-3}$     & 0.3           & 30\,au          \\
		Free-Floating            &  1\,pc            & 1.6  & $10^4$\,M$_\odot$\,pc$^{-3}$  & 0.3        & -             \\
		           
		\hline
	\end{tabular}
\end{table}

We do not include stellar evolution or primordial stellar binaries in the simulations. The stellar masses are sampled from a Maschberger IMF \citep{2013Maschberger}:
\begin{equation} \label{eq:Mimf}
    m(u) = \mu \left[ \left( u \left(G(m_u) - G(m_l)  \right) + G(m_l) \right)^{\frac{1}{1-\beta}} - 1 \right]
           ^{\frac{1}{1-\alpha}},
\end{equation}
where $\mu = 0.2$ M$_\odot$, $u$ is a random number between 0 and 1, $\beta = 1.4$, and $\alpha = 2.3$. $G(m_u)$ and $G(m_l)$ are calculated using Equation \ref{eq:G(m)}, where $m_u$ and $m_l$ correspond to the upper and lower stellar mass limits, respectively:
\begin{equation} \label{eq:G(m)}
    G(m) = \left[ 1 + \left( \frac{m}{\mu} \right)^{1-\alpha}   \right]^{1-\beta}.
\end{equation}
Here, we adopt stellar mass limits of $m_u = 50$ M$_\odot$ and $m_l = 0.1$ M$_\odot$.

Twenty realisations of each set of simulation were run and analysed. Each of these simulations are statistically identical, with the same virial ratio and initial semimajor axis, but with different random number seeds used to initialise the positions, velocities, and masses of the stars.

\subsection{Analysis} 
In each simulation the data is output every 0.01 Myr, which is used to determine the orbital properties of the planetary systems.

\subsubsection{Identification of planetary systems} 
Planetary systems are found by identifying planets and stars that are mutual nearest neighbours and have a negative binding energy i.e. are gravitationally bound.

Triple systems that include a planet are found in a similar way. In this case, all three bodies must be each other's first and second nearest neighbours, the two closest bodies must be gravitationally bound, and the centre of mass of the two closest bodies must be bound to the third body. If a planet is found to be in a triple system, it is logged as such, treated as a binary for the rest of the analysis, then followed up manually if the system is of particular interest.

Planet-planet binaries are found in the same way as star-planet binaries, logged, and followed up manually.

\subsubsection{Classification of orbital type} 
\label{sec:orbit_types}
In any given snapshot a planet is classified as either preserved, captured, stolen, or free-floating.

A planet is classified as free-floating if it is not found to be bound to another star or planet in that snapshot.
A planet that is bound to a star is classified as preserved, captured, or stolen based on its binary history.
A planet is preserved if it is still, and was always, bound to its original star in every prior snapshot.
A planet is captured if it was free-floating in the previous snapshot, but is now bound to a star, whether that be its original star or a different one.
Finally, a planet is classified as stolen if it is bound to a star in the current snapshot, but was bound to a different star in the previous snapshot.
In this way, captured and stolen planets are distinguished based on whether they have been free-floating for a significant period of time.
Since we use 0.01 Myr snapshot intervals, the maximum amount of time that one of our stolen planets could have been free-floating is $<0.01$ Myr.
We believe that this is a reasonable assumption given that this is much shorter than the $\sim0.1$ Myr crossing time of our star-forming regions, and therefore the characteristic timescale over which stellar interactions tend to occur.

The number of each type of planetary orbit is calculated in each snapshot.

\subsubsection{Orbital parameters} 
The semimajor axis and eccentricity are calculated as normal for all planetary systems. The absolute inclination, relative to the coordinate system, is calculated for stolen and captured planets. For preserved planets, the change from its original inclination is used.

\section{Results and Discussion}
\label{sec:ResultsDiscussion}

We first present results for planets in high density ($r=1$ pc, $10^4$ M$_{\odot}$pc$^{-3}$) simulations in Sections \ref{sec:number} to \ref{sec:inclination}, before discussing planets in low density ($r=5$ pc, $100$ M$_{\odot}$pc$^{-3}$) simulations in Section \ref{sec:lowden}.

\subsection{Number of Stolen and Captured Planets}
\label{sec:number} 

\begin{figure*}
    \centering
    \includegraphics[width=\textwidth]{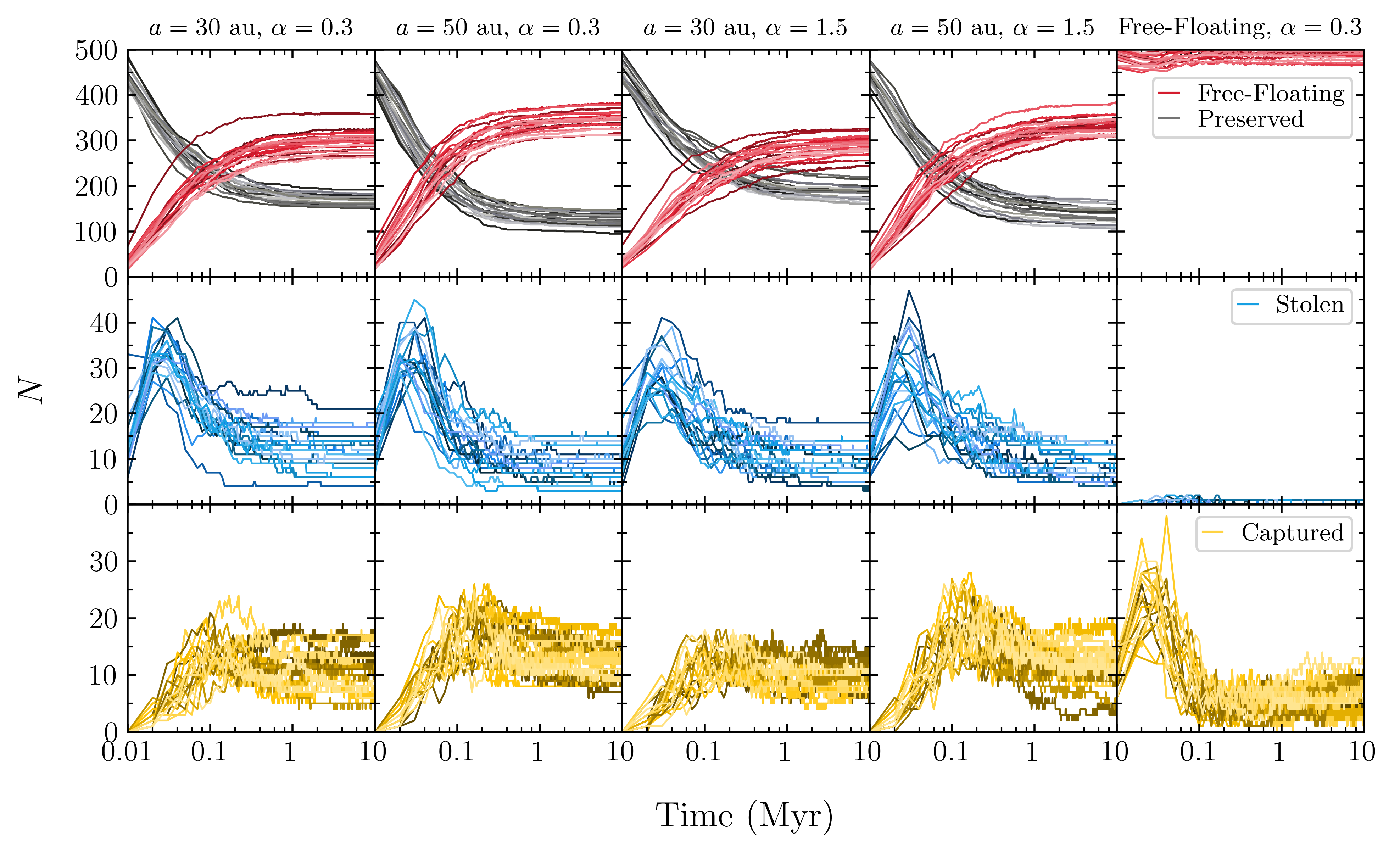}
    \caption{The number of each type of planetary orbit, plotted from the first snapshot at 0.01 Myr to 10 Myr for the higher initial density, $r=1$ pc simulations.
    Each of the first four columns shows results for one of the four bound-planet initial conditions, with lines plotted separately in different shades of the same colour for each of the 20 realisations.
    The final column shows the same for initial conditions where all planets are free-floating with $\alpha=0.3$.
    The number of free-floating, preserved, stolen, and captured planets are shown in shades of red, grey, blue, and yellow, respectively.}
    \label{fig:Freq_main}
\end{figure*}

\subsubsection{Over time}
Most theft and capture happens at early times, when the region is most dense and the stars and planets are more likely to experience encounters. 

For most new planetary systems, the planet is stolen or captured onto an orbit that is either inherently unstable, or is easily destroyed by subsequent interactions. 
In particular, the same high densities that tend to create more stolen and captured systems at early times also increase the likelihood that they will be destroyed.
This leads to the initial peak, which is visible in the stolen and captured planet panels of Figure \ref{fig:Freq_main}, as the the number of stolen and captured planets increase sharply before decreasing and levelling off at around $\approx1$ Myr.
This is the case for every simulation.

After this levelling off at $\approx1$ Myr, an average of $\approx4$\% of planets are either captured or stolen for the remainder of the simulations.
This change over time can be seen in the first four columns of Figure \ref{fig:Freq_main}.

\subsubsection{Effect of initial conditions}
\begin{figure}
    \centering
    \includegraphics[width=\columnwidth]{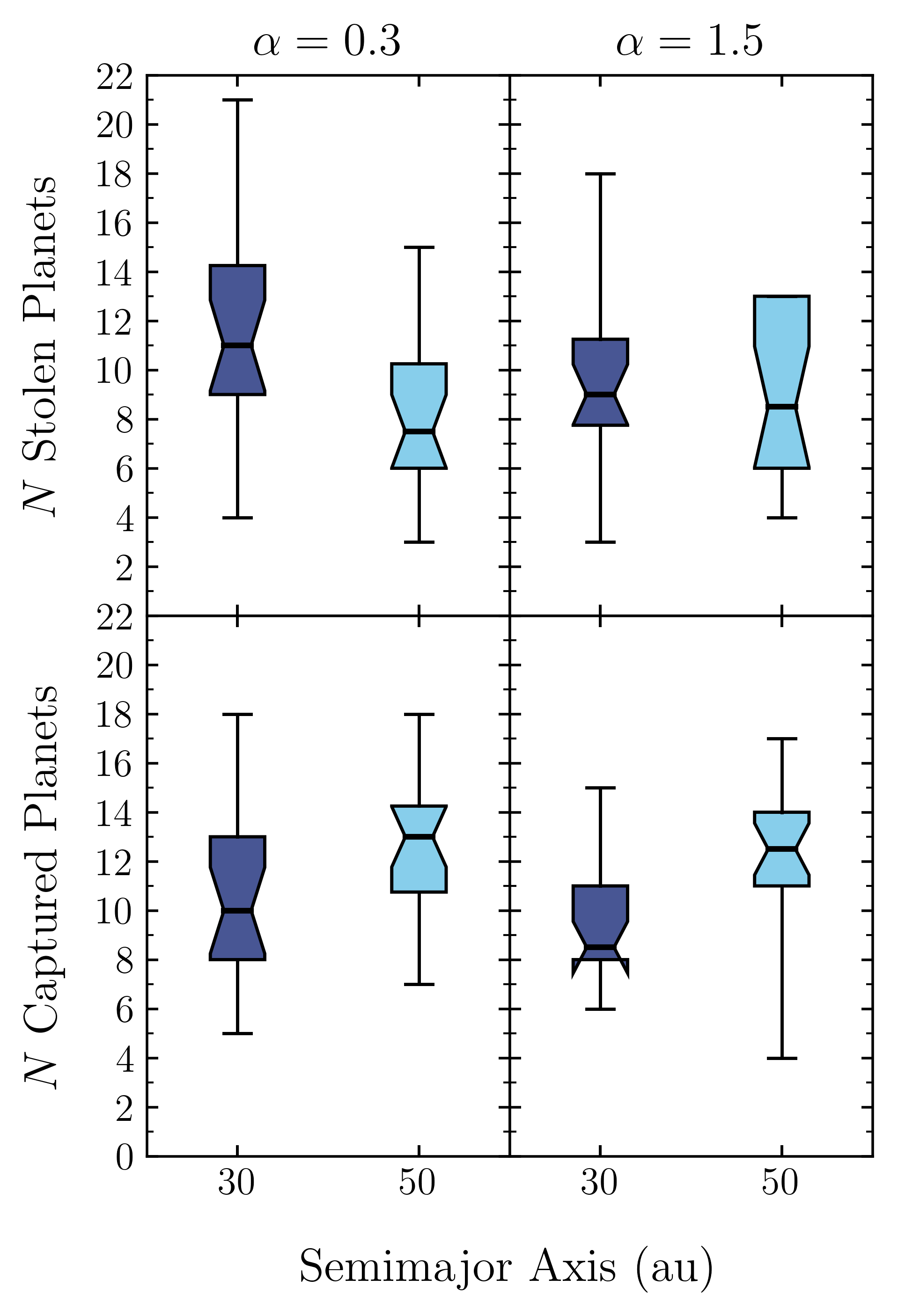}
    \caption{The number of stolen and captured planets at 10 Myr, for each of the 4 initial conditions where the planets are initially bound to stars in $r=1$ pc star-forming regions.
    These results are shown as a separate notched boxplot for each of the initial conditions.
    Each median value is shown as a thick black horizontal line, and the notches are the sloped outer edges which angle outwards from each median line.
    These notches show the 95\% confidence limits for their corresponding median, where any part of the box plot that does not have a sloped outer edge is not within the 95\% confidence interval.
    The second rightmost captured planet box plot has a 95\% confidence limit that is lower than the lower interquartile range.
    This means that the lower notches on this boxplot extend further than the box itself, thereby producing the inverted shape.
    The whiskers show the full range of values.
    There is no upper whisker for the rightmost stolen planet boxplot because the highest value is equal to the upper quartile.
    This is possible for sets of discrete data, and happens here because several of the simulations finish with 13 stolen planets, which is the highest value in that dataset.}
    \label{fig:Freqs_comp}
\end{figure}

The average number of stolen and captured planets at 10 Myr, for simulations where the planets are initially bound, is shown in the boxplots of Figure \ref{fig:Freqs_comp}.
As might be expected, there are on average more captured planets at 10 Myr when the planets are initially placed on the wider 50 au orbits, as shown in Figure \ref{fig:Freqs_comp}.
This is regardless of the initial virial ratio, and is because planets that are initially on wider orbits have a lower binding energy and are more easily ejected from their system during encounters, therefore increasing the number of free-floating planets that are available to be captured.

Figure \ref{fig:Freqs_comp} also compares the final numbers of stolen planets across the four sets of bound-planet initial conditions.
Unlike with captured planets, the average number of stolen planets is higher when they are initially placed at 30 au, rather than 50 au, for a given initial virial ratio.
In fact, for the 30 au initial conditions there are also on average more stolen planets than captured ones.

The notches on each box plot in Figure \ref{fig:Freqs_comp}, which are the sloped outer edges that angle outwards from each median line, represent the 95\% confidence limits of their corresponding median value.
It can be seen from the second panel of Figure \ref{fig:Freqs_comp} that the 95\% confidence limits for the median number of stolen planets fully overlaps for the 30 AU and 50 AU initial semimajor axes.
It should therefore not be concluded that a change in initial semimajor axis affects the final number of stolen planets for the supervirial $\alpha=1.5$ initial conditions.
However, it can be concluded that, for the subvirial $\alpha=0.3$ initial conditions, increasing the initial semimajor axis to 50 AU decreases the final number of stolen planets.
This is the opposite trend than is seen for captured planets, of which there is a higher number for larger initial semimajor axes.

This difference in the final number of stolen planets for the subvirial  ($\alpha = 0.3$) initial conditions is likely caused by the increased number of free-floating planets for the 50 AU initial conditions.
Although the number of stolen planets is consistent at early times for both the $a_p = 50$\,au and $a_p = 30$\,au initial conditions, the medians begin to differ with 95\% confidence after 0.12 Myr - the same time at which the number of stolen planets begins to decrease in all simulations.
It is at this time that the higher number of free-floating planets mean that there are fewer bound planets available to be stolen, and also more free-floating planets which are capable of disrupting a stolen planet's orbit.

This is a clear example of captured and stolen planets being affected in not only different, but in this case opposite ways, by certain initial conditions - underscoring that captured and stolen planets are distinct phenomena that are formed through independent mechanisms.

These results also highlight the significant effect that a planet's initial semimajor axis can have on its fate, in agreement with previous studies \citep{2012ParkerQuanz, 2013HaoEtAl, 2015ZhengEtAl, 2019CaiEtAl}.

\subsubsection{Free-floating initial conditions}
The initial peak in the number of captured planets is larger for the initial conditions where the planets are initially free-floating, rather than bound to a star, as shown in the bottom right panel of Figure \ref{fig:Freq_main}.
This is simply because there are more free-floating planets available to be captured at early times with these initial conditions.

Conversely, however, there are significantly fewer stolen planets in these simulations. The highest number of stolen planets for the free-floating, $\alpha = 0.3$ initial conditions is 2. 
Several of the 20 simulations reach 2 stolen planets within the first several snapshots, but then subsequently drop to 1 or 0.
This is because, in order to be stolen, the planet must first be captured onto an orbit.
In this way, the earliest time at which a planet can be stolen is later for simulations with free-floating initial conditions, thereby reducing the initial peak to only $<2$ stars.
This is an effect that the number of stolen planets does not recover from, as it is at these earlier times when planetary interactions, including theft, are more likely.

This has implications for studies that use free-floating initial conditions to investigate planet capture and theft. 
For example, \citet{2017ParkerEtAl} used $N$-body simulations with free-floating initial conditions to investigate the origin of Planet 9 in the context of it having formed around a star other than the Sun.
The choice of free-floating initial conditions will have reduced the frequency of planet theft, thereby also reducing the total number of planets which are orbiting a new star at the end of the simulations.
It may therefore have affected how their conclusion, that the likelihood of Planet 9 having originated from another planetary system is almost zero, compares to that of other similar similar studies such as \citet{2016MustillEtAl} and \citet{2016LiAdams}, who find the chance of Planet 9 having formed via theft to be non-zero.
Although, as discussed in \S\ref{sec:a} and shown in Figures \ref{fig:Cfreq_A} and \ref{fig:AvsE}, we find that planets with semimajor axes in Planet 9's predicted orbital range are most likely to have been captured. 

Nevertheless, it is important to be aware that free-floating initial conditions suppress the formation of stolen planetary systems, compared to corresponding simulations in which the planets are initially bound.

We do not discuss free-floating initial conditions further, due to the low number of stolen planets. All further discussion relates to the sets of simulations where the planets are initially bound to host stars.

\subsubsection{A note on planet-planet systems}
We identify 5 planet-planet systems across all of our simulations.
These planets tend to have very rich dynamical histories.
For example, one planet in the $\alpha=0.3$, $a_p = 30$\,au simulations has repeated interactions with another star and planet. 
This leads to it being in a temporary planet-planet system.
It is removed and then recaptured by its original star throughout the simulation, and ends the 10 Myr as free-floating, after becoming unbound from its original star a final time.

All of these planet-planet systems are short-lived, and are each only identified in one 0.01 Myr snapshot.
Such systems would therefore be very unlikely to be observed, and we do not include them in any formal analysis.

\subsection{Semimajor Axis}
\label{sec:a}

\begin{figure}
    \centering
    \includegraphics[width=\columnwidth]{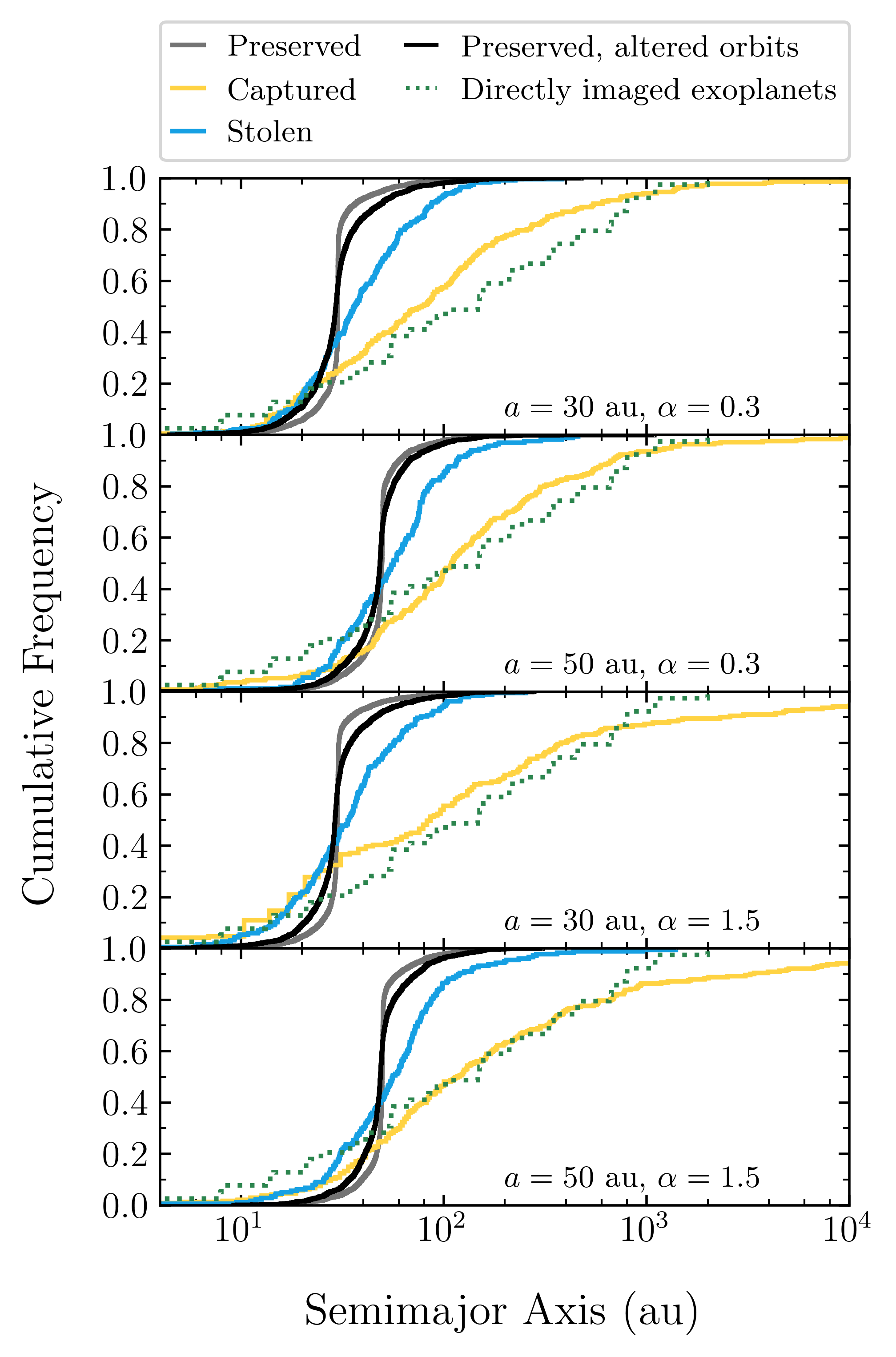}
    \caption{Semimajor axis distribution for planets that are bound to a star after 10 Myr, categorised according to the three types of planetary orbit: preserved (grey), captured (yellow), and stolen  (blue). We also show the distributions for preserved planets with altered orbits (the black lines), defined as a change in eccentricity of more than $\Delta e = 0.1$ and/or a change in semimajor axis of $\pm$10\,per cent. Results are summed and shown for all 20 realisations of all 4 sets of bound-planet, $r=1$ pc initial conditions, which are shown in separate panels.
The semimajor axis distribution for directly detected exoplanets is also shown for comparison as a dotted green line in each panel.
For the observed exoplanets, the semimajor axis is used where the data is available, otherwise the projected separation is plotted.}
 \label{fig:Cfreq_A}
\end{figure}

\begin{figure*}
    \centering
    \includegraphics[width=\textwidth]{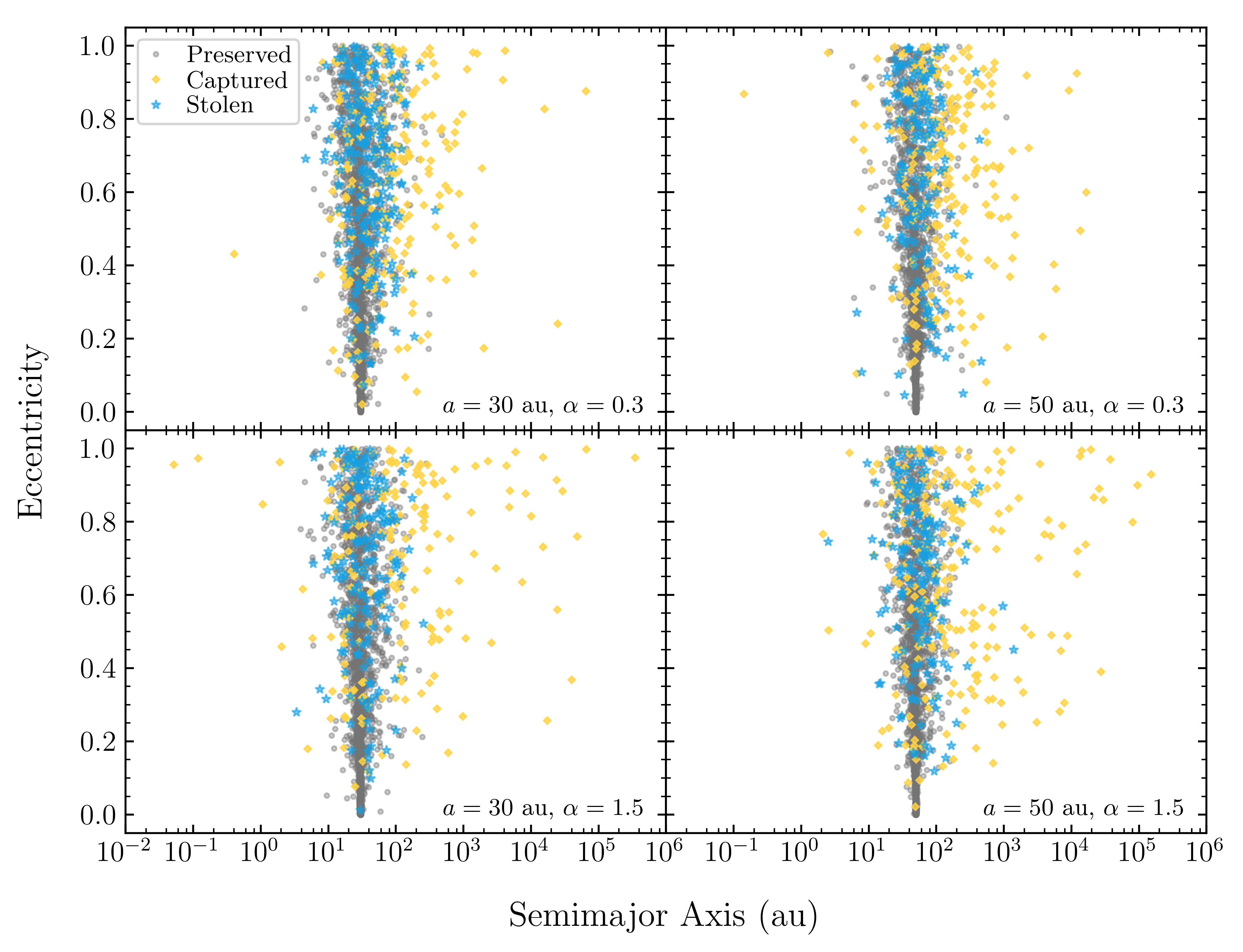}
    \caption{Characteristic `firework' plots of the semimajor axis vs eccentricity distribution. Results are shown for planets that are bound to a star after 10 Myr, divided into the three types of planetary orbit: preserved (grey), captured (yellow), and stolen (blue). Each planet is shown as a semi-transparent point. Results are summed and shown for all 20 realisations of all 4 sets of bound-planet, $r=1$ pc initial conditions, which are shown in separate panels.}
    \label{fig:AvsE}
\end{figure*}

The semimajor axis distributions for preserved, captured, and stolen planets differ significantly, as shown in Figures \ref{fig:Cfreq_A} and \ref{fig:AvsE}.
Of the planetary systems that are bound at 10 Myr, Figure \ref{fig:Cfreq_A} shows that $\sim20\%$ of preserved planets have their semimajor axes disrupted such that it is greater than the initial 30-50 au value.
In contrast, between between $\sim60-80 \%$ of captured and stolen planets have semimajor axes greater than this.

One of the most noticeable trends in Figures \ref{fig:Cfreq_A} and \ref{fig:AvsE} is that captured planets tend to be on wider orbits than stolen and preserved planets.
This is the case for all initial conditions and for both the average and upper limit values of $a$.
This is a difference that could be seen observationally and used to distinguish a population of captured planets, as these results suggest that an exoplanet with an observed semimajor axis of $\gtrsim500$ AU has been captured.
Comparing this to Planets 9's predicted orbital range of $a \sim 400-800$ AU), eccentricity ($e \sim 0.2-0.5$), and inclination ($i \sim 15-25 ^{\circ}$) \citep{2019BatyginEtAl, 2020FiengaEtAl}, these results suggest that, should Planet 9 exist, it is most likely to have been captured.

Captured planets can have these wider orbits because capture tends to happen when the planet and star exit the cluster at the same time and in the same direction \citep{2012PeretsKouwenhoven, 2014ParkerMeyer}.
In these cases, the star and planet will tend to be relatively isolated, without other gravitational interactions that may interfere with the new orbit.
This means that a more loosely bound orbit will tend to remain bound for longer.

The semimajor axis distribution of stolen planets is similar to the semimajor axis distribution of preserved planets.
It is possible that using a realistic range of initial semimajor axes would add spread to the distributions, thereby causing them to be more similar, and harder to separate.
However, comparing previous studies shows that results are consistent between simulations that use single semimajor axis values, and those that either select the semimajor axes from a distribution or use a wide range.
For example, results are consistent when a single semimajor axis of 30 AU is used, and when a range of semimajor axes that have been sampled from a distribution that has a median of 30 AU \citep[e.g.][]{2012ParkerQuanz, 2015ForganEtAl, 2015ZhengEtAl}.

Many of the preserved planets have relatively unaltered orbital properties, and as we used a delta function as our initial semimajor axis (all have $a = 30$\,au or $a = 50\,au$, and $e = 0$), the cumulative distribution of the preserved planets (shown in grey) may be dominated by planets that have not experienced an interaction with a passing star(s). 

To check this, we plot a second (black) line for preserved planets, but limit this to systems whose orbits are altered (defined as the semimajor axis changing by $\pm$10\%, and/or the eccentricity changing by $\Delta e \geq 0.1$, following \citet{2012ParkerQuanz})\footnote{In Appendix~\ref{appendix} we show the results when assuming more drastic properties for the altered orbits, i.e.\,\,the semimajor axis changing by $\pm$50\%, and/or the eccentricity changing by $\Delta e \geq 0.5$.}. Whilst the altered preserved planets (the black line) display a distribution that is closer to the stolen planets (the blue line), it is still much closer to the distribution for all preserved systems (the grey line), suggesting that the distributions of preserved and stolen planets are indeed distinct and different.

Figure \ref{fig:Cfreq_A} also shows the semimajor axis distribution for directly imaged exoplanets\footnote{This data was taken from the NASA Exoplanet Archive on 18/02/2021.}
(in most cases the projected separation data is used in place of the semimajor axis). We emphasise that the sample of directly imaged exoplanets is likely to be incomplete and affected by statistical biases, and we have no information on whether each planet in the sample formed at its observed separation, or whether some process(es) moved the planets.  

Despite these caveats we note that there is general agreement between our captured planet semimajor axis distribution and that of directly imaged exoplanets.
This is the case for all of the high density initial conditions shown in Figure \ref{fig:Cfreq_A} suggesting that, if directly imaged planets on wide orbits are a result of dynamics, they are likely to be captured rather than stolen or preserved. If this is the case, then we might expect to discover more such planets on extremely wide ($>$1000\,au) orbits, which are present in our simulation data but not yet in the observational data \citep[though see][for efforts towards this parameter space]{Durkan2016}. \\
\\

\subsection{Eccentricity} 

\begin{figure}
    \centering
    \includegraphics[width=\columnwidth]{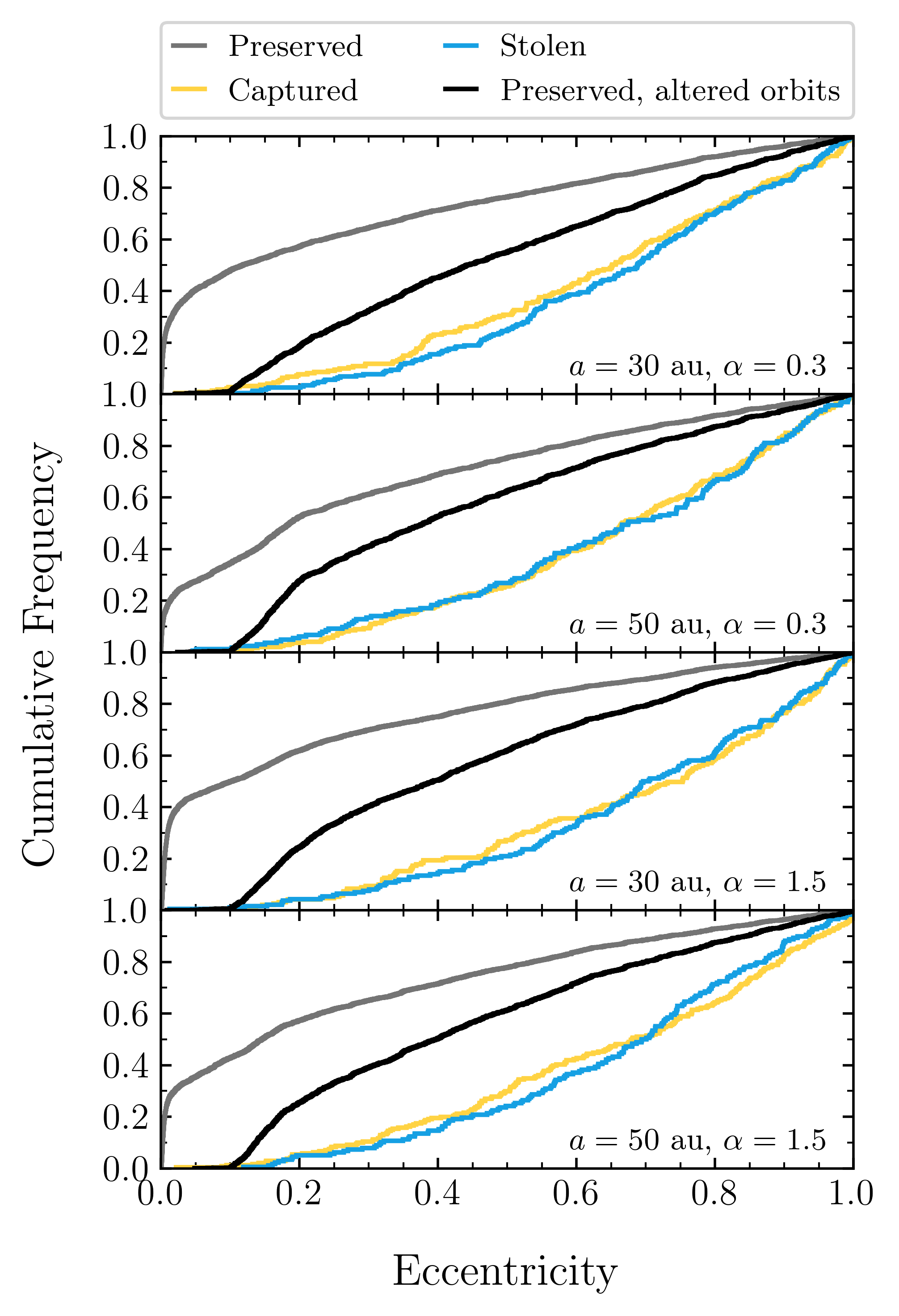}
    \caption{Eccentricity distribution for planets that are bound to a star after 10 Myr, categorised according to the three types of planetary orbit: preserved (grey), captured (yellow), and stolen (blue). We also show the distributions for preserved planets with altered orbits (the black lines), defined as a change in eccentricity of more than $\Delta e = 0.1$ and/or a change in semimajor axis of $\pm$10\,per cent. The results are summed and shown for all 20 realisations of all 4 sets of bound-planet, $r=1$ pc initial conditions, which are shown in separate panels.}
    \label{fig:Cfreq_E}
\end{figure}

Figure \ref{fig:Cfreq_E} shows that the eccentricity distribution for captured and stolen planets is thermal. 
This is expected for captured planets as a thermal distribution is seen for binary systems that have formed dynamically \citep{1975Heggie}.

The eccentricity distribution for the preserved planets (the grey line) is very different to those for the captured and stolen planets (the yellow and blue lines, respectively). The distribution of eccentricities for the preserved planets could be dominated by the systems that have not experienced an encounter, which still have eccentricity values around zero. However, if we again use the definition from \cite{2012ParkerQuanz}, that a planet is disrupted if its semimajor axis is changed by 10\% or its eccentricity is increased above 0.1, we find that where $\sim20 \%$ of preserved planets have their orbits disrupted in terms of semimajor axis, $\sim~40-50 \%$ are disrupted in terms of eccentricity.

If we plot the eccentricity distributions of preserved planets with altered orbits (the black line in Fig.~\ref{fig:Cfreq_E}), we can see that they straddle the parameter space between the entire preserved planet population, and the captured and stolen planets. However, this is mainly due to our definition for when an orbit is altered, i.e.\,\,$e>0.1$. The \emph{shape} of the distribution is still markedly different to the distributions of the captured and stolen planets.
(Inspection of Figure \ref{fig:AvsE} shows that it is also possible for some stolen and captured planets to have semimajor axes that are indistinguishable from preserved planets that have not had their orbit disrupted.)

The results of K-S tests confirm that the eccentricity distributions for captured and stolen planets are all very similar (K-S statistics $<0.1$), to a relatively high confidence (p-values ranging from 0.33 to 0.80).
This is the case for all sets of bound-planet, $r=1$ pc initial conditions.
The null hypothesis, that the eccentricity distributions for captured and stolen planets are sampled from the same underlying distribution, therefore cannot be rejected for these simulations.
Eccentricity can therefore not be used to distinguish captured from stolen planets in observations of exoplanets that have likely formed in dense regions.

The eccentricity distribution could, however, be useful for distinguishing the planets in a population that are preserved from those that are new systems, which have formed dynamically through either capture or theft.

\subsection{Inclination} 
\label{sec:inclination}

\begin{figure}
    \centering
    \includegraphics[width=\columnwidth]{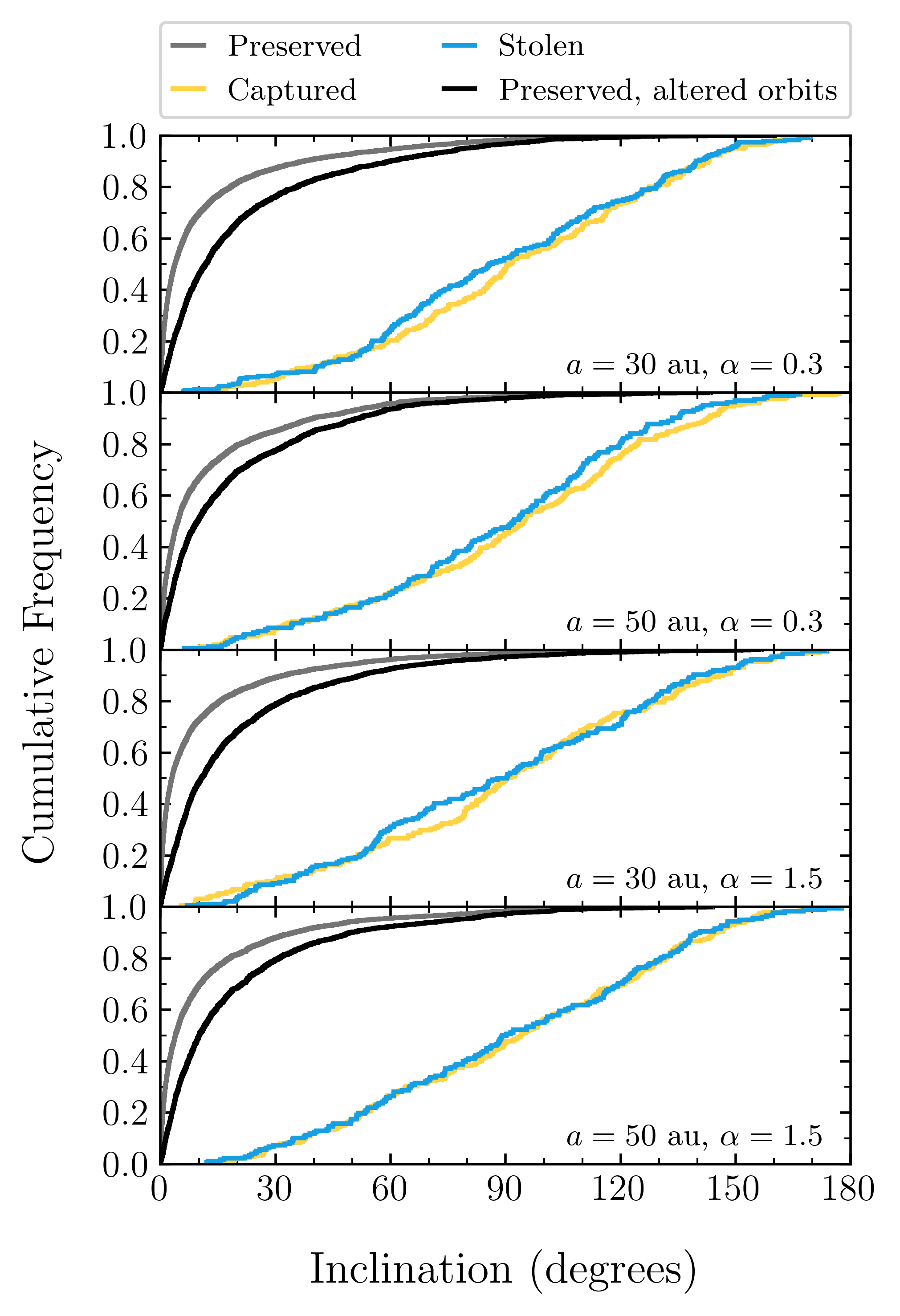}
    \caption{Inclination distribution for planets that are bound to a star after 10 Myr, categorised according to the three types of planetary orbit: preserved (grey), captured (yellow), and stolen (blue). We also show the distributions for preserved planets with altered orbits (the black lines), defined as a change in eccentricity of more than $\Delta e = 0.1$ and/or a change in semimajor axis of $\pm$10\,per cent. The results are summed and shown for all 20 realisations of all 4 sets of bound-planet, $r=1$ pc initial conditions, which are shown in separate panels.}
    \label{fig:Cfreq_I}
\end{figure}

In Fig.~\ref{fig:Cfreq_I} we plot the distribution of planets' inclinations for preserved (grey lines), captured (yellow lines) and stolen (blue lines) planets. We also plot the preserved planets whose orbits have been significantly altered (defined as a change in eccentricity of more than $\Delta e = 0.1$ and/or a change in semimajor axis of $\pm$10\,per cent, the grey lines). For the captured and stolen planets, we assume the inclination with respect to the plane, and for the preserved planets we plot the relative change in inclination (as the initial inclination angles were randomly chosen). Figure \ref{fig:Cfreq_I} shows that, as expected, planets are stolen and captured onto orbits with random inclinations. 

In our simulations, we find $\sim20\,\%$ of preserved planets have their inclinations disrupted by more than $10 \%$. This is comparable to the percentage that are disrupted in terms of their semimajor axis. As would be expected, this shows that, should a planet be observed in a system with a disk or other planets, the relative inclination could be used to determine whether the exoplanet has likely been captured or stolen from another star \citep[as an interaction that would disrupt one planet would likely induce instabilities in the orbits of the other planets, e.g.][]{Malmberg2007} -- a small relative inclination might imply the planets had formed in the same system.

\subsection{Effects of Lower Density}
\label{sec:lowden}
\begin{figure}
    \centering
    \includegraphics[scale =0.85]{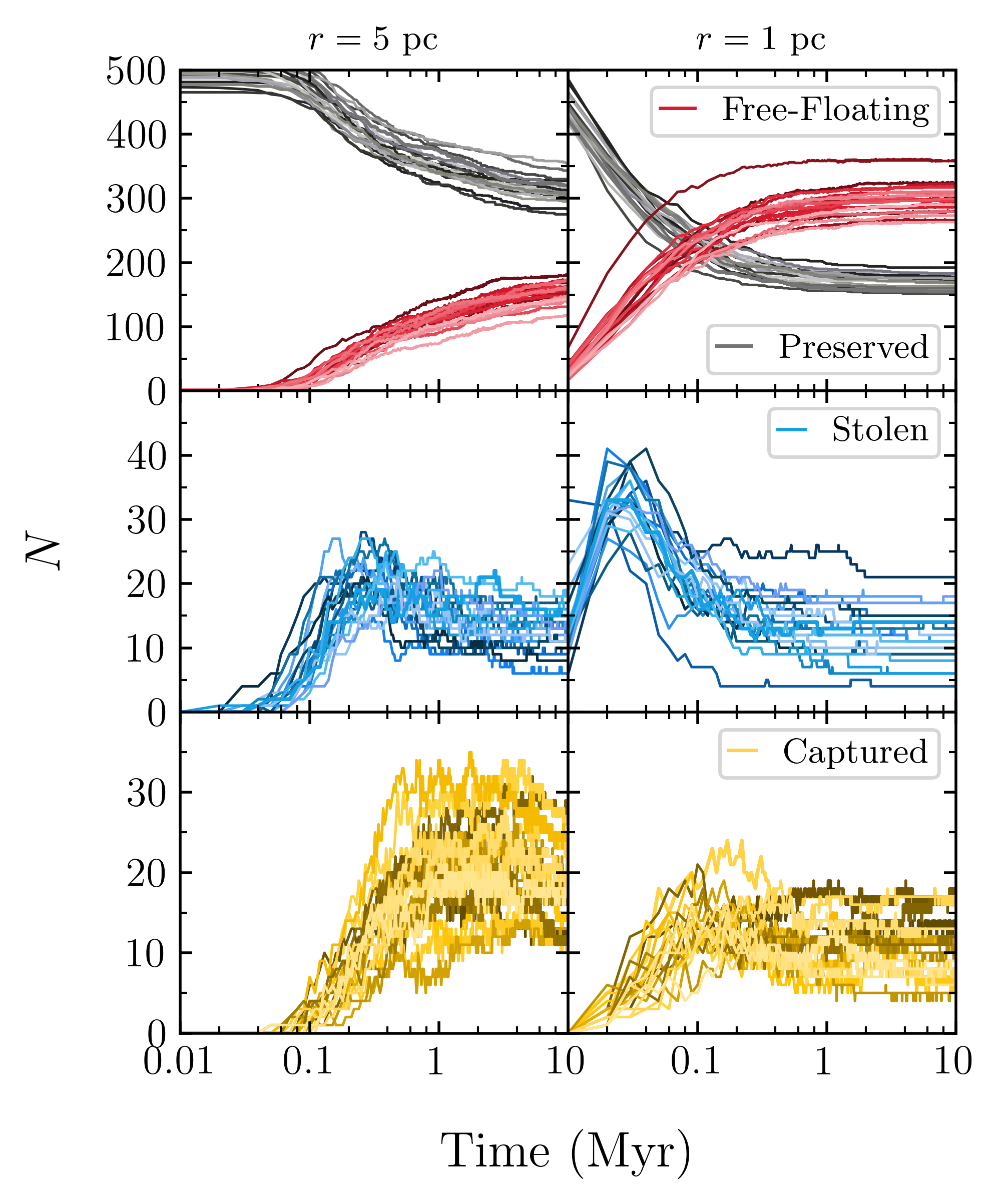}
    \caption{The number of each type of planetary orbit over 10 Myr, plotted from the first snapshot at 0.01 Myr to 10 Myr.
The first column shows this for the lower density, $r=5$ pc, $\alpha=0.3$, $a_p = 30$\,au initial conditions.
The second columns shows this for the higher density $r=1$ pc, $\alpha=0.3$, $a_p = 30$\,au initial conditions.
    The number of free-floating, preserved, stolen, and captured planets are shown in shades of red, grey, blue, and yellow, respectively. Each individual simulation is shown in a different hue of the same colour.}
    \label{fig:Freq_LowDen}
\end{figure}
\begin{figure*}
    \centering
    \includegraphics[width=\textwidth]{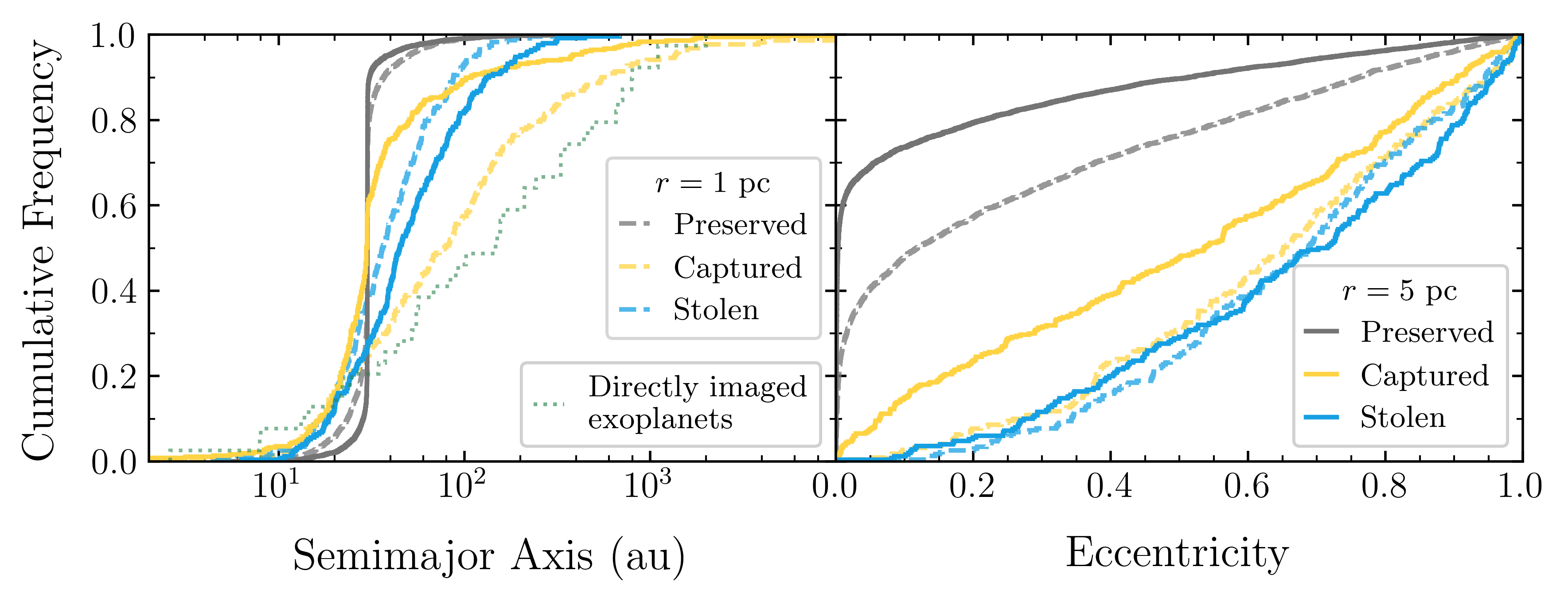}
    \caption{Semimajor axis and eccentricity distribution for planets that are bound to a star after 10 Myr, categorised according to the three types of planetary orbit: preserved (grey), captured (yellow), and stolen (blue). 
Results are summed and shown for all of the lower density $r=5$ pc, $\alpha=0.3$, $a_p = 30$\,au initial conditions and the higher density $r=1$ pc, $\alpha=0.3$, $a_p = 30$\,au initial conditions.
The lower density results are shown as solid lines, and the higher density results are shown as slightly transparent dashed lines. The semimajor axis distribution for directly detected exoplanets is also shown for comparison as a dotted green line in the lefthand panel.}
    \label{fig:LowDensity_CFreq}
\end{figure*}

For comparison, we have also run a set of simulations using lower density  ($100$ M$_{\odot}$pc$^{-3}$) initial conditions, with an initial radius of 5 pc. 
We compare these lower density $\alpha=0.3$, 30 au simulations to the higher density $\alpha=0.3$, 30 au simulations in Figures \ref{fig:Freq_LowDen} and \ref{fig:LowDensity_CFreq}. 

Figure \ref{fig:Freq_LowDen} compares the number of each type of planetary orbit over time, in the same way as Figure \ref{fig:Freq_main}.
The first difference that can be seen is that the low density initial conditions produce fewer free-floating planets than preserved planets after 10 Myr.
This is in contrast to the high density initial conditions which produce more free-floating planets than preserved planets, as more are ejected from their birth system.

It can be seen from the bottom panels of Figure \ref{fig:Freq_LowDen} that the lower density initial conditions lead to more captured planets after 10 Myr.
This is because, in this less extreme environment, although there are fewer free-floating planets available to be captured, captured systems that do form are more likely to survive. 
There no significant change to the average number of stolen planets when the density is lowered.
However, there is a smaller spread in the number of stolen planets for the low density initial conditions.

Both the semimajor axis and eccentricity distributions are affected by the initial density being lowered. This is shown in Figure \ref{fig:LowDensity_CFreq}, where the higher density results are shown as slightly faded dashed lines, and the low density results are shown as solid lines.
More stolen planets are able to stay stable on wider orbits for the lower density initial conditions, therefore broadening the stolen planet semimajor axis distribution.
For captured planets, the top $\sim5\%$ of the semimajor axis distribution is similar, regardless of density.
However, the additional numbers of planets that are captured in the lower density simulations tend to preferentially fill the center of the distribution.
This has the effect of making the captured planet semimajor axis distribution more similar to that of the stolen and preserved planets.
Nevertheless it is still the case that a planet on an orbit wider than $\sim500$ au is most likely to be captured, regardless of the initial density of the star-forming region.

The eccentricity distribution of the stolen planets remains thermal, as shown in the right panel of  Figure \ref{fig:LowDensity_CFreq}.
However, the eccentricity distribution of the captured planets flattens for the low density simulations.
This means that, at these low densities, the eccentricity distribution of the stolen, captured, and preserved planets are all distinct from each other.
Eccentricity data could therefore be used to separate populations of stolen, captured, and preserved planets from each other in the exoplanet data, should an estimate of their formation density be available (see \citealt{2020WinterEtAl, 2021AdibekyanEtAl}).

\subsection{Effects of Substructure}
\label{sec:substruct}

We now examine the effects of the initial degree of substructure in the simulations. We do this by keeping the initial stellar density constant, and then compare the frequencies of different types of planetary systems with differing amounts of substructure.

\begin{figure*}
    \centering
    \includegraphics[scale =1.05]{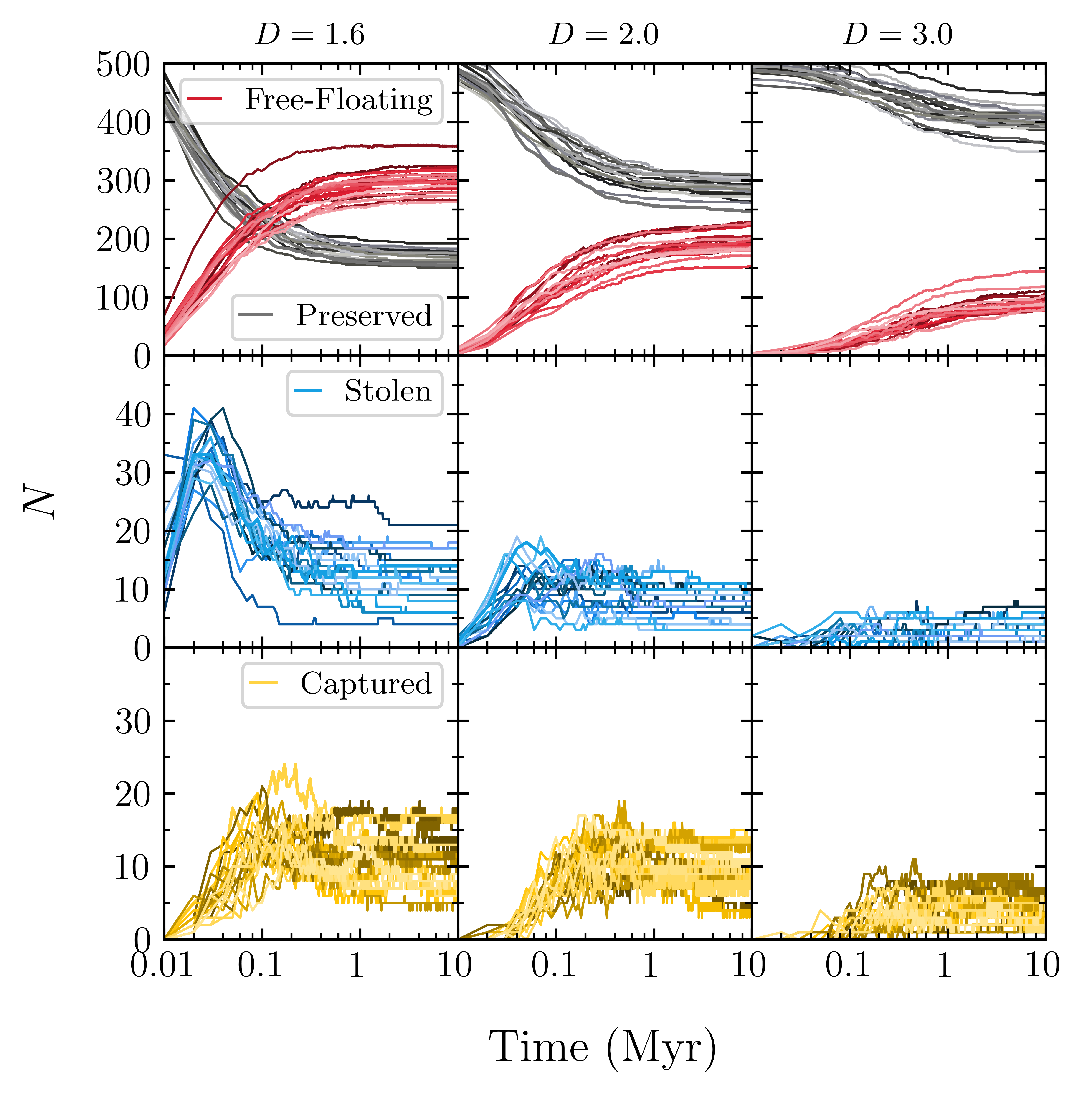}
    \caption{The effect of varying the initial degree of substructure on the number of each type of planetary orbit over 10 Myr, plotted from the first snapshot at 0.01 Myr to 10 Myr.
      The first column shows this for simulations with a high amount of initial substructure, $D=1.6$, with $\alpha=0.3$, $a_p = 30$\,au as the initial conditions (our `default' simulations).
The second column shows this for the simulations that have a moderate degree of substructure ($D=2.0$), with $r=0.5$ pc, $\alpha=0.3$, $a_p = 30$\,au as the initial conditions. The third column shows the results for simulations where their is no initial substructure ($D=3.0$), with  with $r=0.25$ pc, $\alpha=0.3$, $a_p = 30$\,au as the initial conditions. The radii are set such that the initial stellar densities are all $10^4$\,M$_\odot$\,pc$^{-3}$ in the three sets of simulations. 
    The number of free-floating, preserved, stolen, and captured planets are shown  in shades of red, grey, blue, and yellow, respectively. Each individual simulation is shown in a different hue of the same colour.}
    \label{fig:substructure_freqs}
\end{figure*}

  In Fig.~\ref{fig:substructure_freqs} we show the numbers of preserved, stolen and captured planets, as well as the numbers of free-floating planets. Changing the fractal dimension has  a significant effect of the results. The main result is that the number of stolen and captured planets is higher in the more substructured simulations, and the number of free-floating planets is also much higher (conversely, in the more substructured simulations, the number of preserved planets is lower).

  All this is despite the stellar densities in these simulations being identical ($\tilde{\rho} = 10^4$\,M$_\odot$\,pc$^{-3}$). In the non-substructured simulations $D = 3.0$), the stellar density actually remains higher throughout the duration of the simulation, so the differences must be due to the evolution of the clumps of substructure. 

  The more substructured simulations have more velocity correlation in the clumps of substructure \citep{2004GoodwinWhitworth,2020DaffernPowellParker}, which facilitates a higher degree of violent relaxation \citep{LyndenBell67}, leading to the collapse of the clumps of substructure. This process enhances the number of exchange interactions, as well as increasing the number of systems that break apart. The more correlated velocities in the substructured simulations also facilitate a higher rate of capture \citep{Kouwenhoven10,2012PeretsKouwenhoven}.
 
\begin{figure*}
    \centering
    \includegraphics[scale=0.85]{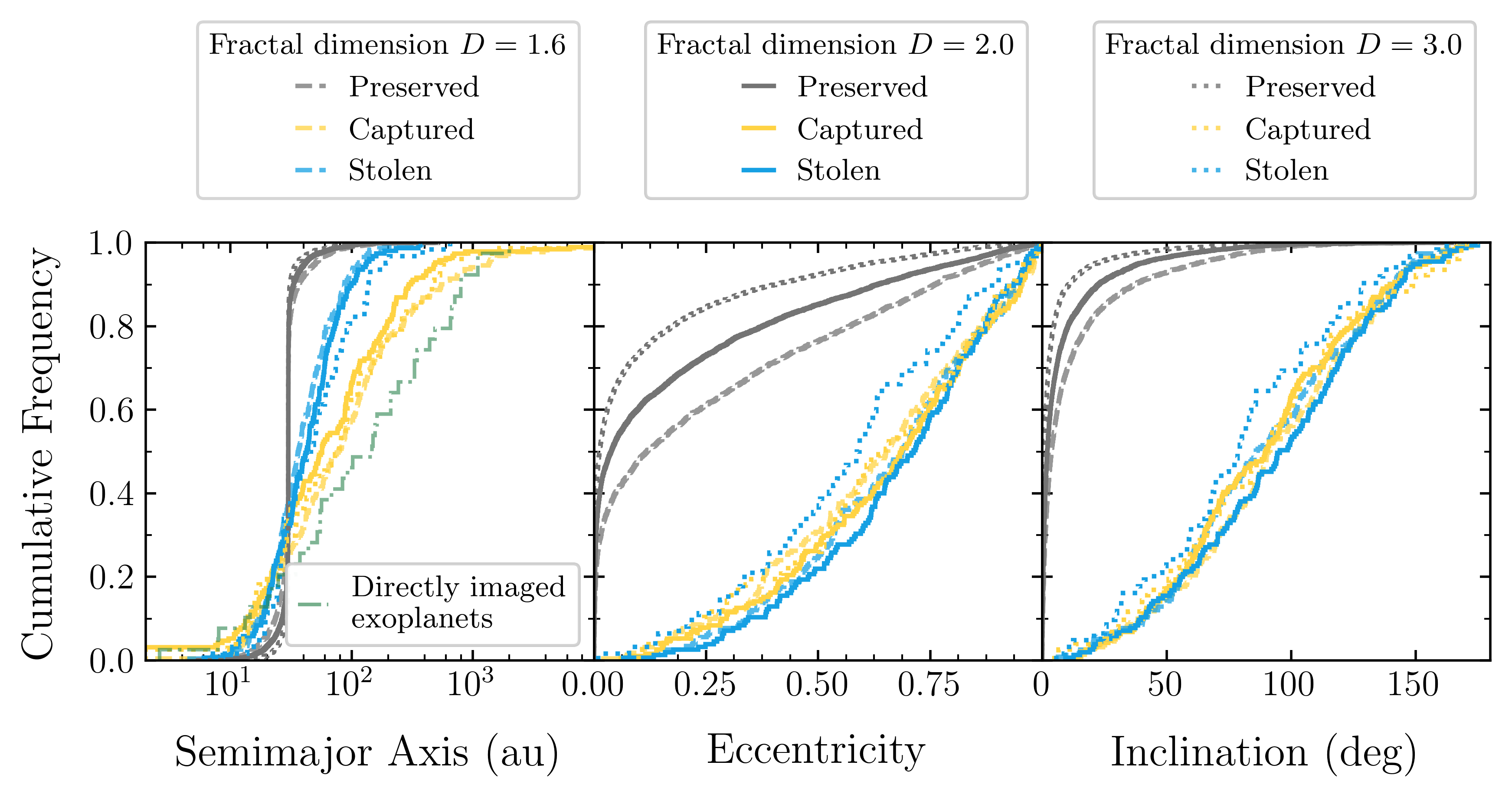}
    \caption{Semimajor axis, eccentricity and inclination distributions for planets that are bound to a star after 10 Myr, categorised according to the three types of planetary orbit: preserved (grey), captured  (yellow), and stolen (blue). 
      Results are summed and shown for all simulations with differing amounts of substructure; those with a high degree of spatial and kinematic structure ($D=1.6$) are shown by the dashed lines, those with a moderate degree of substructure ($D=2.0$) are shown by the solid lines, and those with no substructure ($D=3.0$) are shown by the dotted lines. The semimajor axis distribution for directly detected exoplanets is also shown for comparison as a dot-dashed green line in the lefthand panel. }
     \label{fig:substructure_orbits}
    \end{figure*}

Despite the significant differences in the numbers of preserved, free-floating, captured and stolen planets as a function of the amount of substructure, the distributions of the orbital parameters (semimajor axis, eccentricity and inclination) are relatively constant for the different substructure regimes, as shown in Fig.~\ref{fig:substructure_orbits}.

\subsection{Comparison to Previous Work}
As has already been highlighted by other studies \citep[e.g. that of][when comparing their results to Parker \& Quanz 2012]{2013CraigKrumholz}, differences in the initial conditions and method of simulation can result in large differences in the number of free-floating planets, as well as the numbers of stolen, captured, and preserved planets.

In terms of the number of free-floating planets, we find that $\gtrsim$50\% of planets are free-floating at the end of these simulations, depending on the initial conditions.
This means that they are of order 10 times more common than stolen and captured planets, and $\sim 1.5 - 2$ times more common than planets that remain bound to their original star.
This is higher than the $~10$\% obtained by \citet{2012ParkerQuanz} for simulations where planets are placed at 30 au, and is likely caused by our use of a fractal dimension of 1.6, rather than 2.
This leads to initial densities that are $\sim10$ times higher in our high density, $r=1$ pc simulations (of order $10^4$\,M$_\odot$\,pc$^{-3}$, compared to $10^3$\,M$_\odot$\,pc$^{-3}$, in \citealt{2012ParkerQuanz}).
Although it is unclear whether many star-forming regions have initial densities of this magnitude \citep{2014Parker}, the purpose of this paper is primarily to investigate the orbital properties of stolen and captured planets in the most extreme star-forming environments.

In terms of the frequencies of stolen and captured planets, our $\approx4$\% is an order of magnitude higher than the 0.4\% obtained by \citet{2012ParkerQuanz}, due to their $N$-body simulations having initial densities that are an order of magnitude lower than those used in this paper.

Although the frequencies of each type of planet can vary significantly in this way, the orbital ranges and distributions in $a-e$ space are broadly consistent with previous studies. 

As noted in Section~1, the alternative approach in which planetary simulations are evolved separately but within global simulations of star-forming regions is limited by the fact that planets cannot be come free-floating in the star-forming region (and possibly then (re)captured), nor can these simulations model exchange interactions where planets can be stolen from other stars. As such, a detailed comparison with these studies is not possible, and would also be hamstrung by the lack of long-term evolution of the planets in our simulations. This is a major shortcoming of our approach of modelling the planets within the star-forming regions via a `brute force' method.

\section{Conclusions} 
\label{sec:Conlcusion}
We use $N$-body simulations of planets in star-forming regions to investigate the dynamical evolution of the planets within them.

Our star-forming regions are highly substructured initially, with a fractal dimension of $D=1.6$, and we model star-forming regions with both sub and supervirial initial conditions (virial ratios of 0.3 and 1.5 respectively).
There are 1000 stars in each star-forming region, half of which have planets initially placed at either 30 AU or 50 AU.

The dynamical evolution of the star-forming regions is followed for 10 Myr, and we focus on the orbital properties of stolen planets (that have been directly exchanged between stars during an encounter), and how these compare to that of captured free-floating planets and planets that remain bound to their original star.

\medskip
\noindent Our main results are summarised as follows:
\begin{enumerate}
    \item We find that planet theft and capture should be seen as two distinct mechanisms, and should therefore be treated and analysed as such wherever feasible.
    Our evidence for this is twofold.
    First, we find that the number of stolen and captured planets is a strong function of the initial conditions.
    Second, we find that the orbital distributions of stolen and captured planets are distinct.
    The evidence of these differences is lost when stolen planets are categorised together with captured ones.
    
    \item The orbital properties of stolen, captured, and preserved planets are distinct enough that these characteristics could be used to distinguish their formation channel if an estimate of the initial conditions of their star-forming region are known.
    Although there is overlap between orbital property distributions, combining semimajor axis and eccentricity data could observationally distinguish populations of stolen, captured, and preserved planets that were born in high density star-forming regions.
    For planets that have formed in such regions, the semimajor axis distribution can separate captured planets from those that are stolen and preserved, and eccentricity can then separate stolen planets from those that are preserved.
    If available, inclination data would also be useful in distinguishing planets that have been captured or stolen from another system.
    This analysis could be performed on populations of planets as a whole to estimate the frequency of each type of planet, or on individual systems to estimate the probability that a planet is stolen, captured, or has remained orbiting in its original system.

	\item Regardless of the initial conditions, we find that a planet with a semimajor axis of $\gtrsim 500$ au is mostly likely a captured planet.
	And comparing our orbital property results to Planet 9's predicted orbital range of $a \sim 400-800$ AU and $e \sim 0.2-0.5$ suggests that, should it exist, Planet 9 is most likely to have been captured, rather than stolen.

	\item The semimajor axis distribution of our captured planets is in some instances similar to the semimajor axis distribution of exoplanets found by direct imaging (although these data are likely to be biased and incomplete). There is debate about the formation mechanism of exoplanets, especially those on wide orbits, and our results suggest that they may be captured, formerly free-floating planets, as has been suggested by previous studies \citep{2012PeretsKouwenhoven}.
    
    \item We find that theft and capture are relatively common, with $\sim2\%$ of planets being in stolen systems, and $\sim2\%$ in captured systems at 10 Myr.
    The likelihood of planet theft and capture should therefore not be seen as negligible, especially for star-forming regions which may have had relatively dense initial conditions, as simulated here.

	\item Smaller (30 au) initial semimajor axes lead to more stolen planets than captured ones after 10 Myr, whilst larger (50 au) initial semimajor axes lead to more captured planets than stolen ones. This suggests that the outcome of dynamical interactions has a strong dependance on the planet's semimajor axis.
    
        \item In simulations where all of the planets are initially free-floating, rather than bound to a star, the incidence of planet theft is negligible. This can give the false impression that planet theft is very rare, and negligible compared to capture, when this is not the case in simulations where the planets are all initially bound to stars.
          \item The initial degree of spatial and kinematic substructure in a star-forming region is as important as the stellar density parameter in determining how many, and the extent to which, planetary systems are disrupted. Lower densities result in fewer free-floating planets, and fewer stolen planets, but increase the likelihood of planetary capture. Simulations with more substructure lead to more stolen planets, more captured planets, and more free-floating planets (fewer preserved planets), even when the stellar densities are identical. This is caused by the violent relaxation within the clumps of substructure. Varying other parameters, such as the initial virial ratio, have much more modest effects.
\end{enumerate}

In general, comparing our results to that of other papers illustrates the significant effects that initial conditions can have on young planetary systems. These interactions can be not only destructive, but also lead to the creation of new and unique planetary systems.

\section*{Acknowledgements} 
We thank the anonymous referee for their suggestions and constructive criticism of our manuscript. ECD acknowledges support from the UK Science and Technology Facilities Council in the form of a
PhD studentship. RJP acknowledges support from the Royal Society in the form of a Dorothy Hodgkin Fellowship. Part of this work has been carried out within the framework of the National Centre of Competence in Research ``PlanetS'' supported by the Swiss National Science Foundation. SPQ acknowledges the financial support of the SNSF. For the purpose of open access, the authors have applied a Creative Commons Attribution (CC BY) license to any Author Accepted manuscript version arising. 

\section*{Data Availability}
The data underlying this article will be shared on reasonable request to the corresponding author.




\bibliographystyle{mnras}
\bibliography{references} 


\appendix

\section{Preserved planets with altered orbits}
\label{appendix}

In our analysis we have differentiated between all planets that are still orbiting their parent star (`preserved') and planets orbiting their parent stars whose orbits have been significantly altered. In the main part of the paper, we deem a planet's orbit to be altered if the eccentricity changes by more than 0.1, and/or the semimajor axis changes by $\pm$10\,per cent. 

In this Appendix, we repeat Figures~\ref{fig:Cfreq_A},~\ref{fig:Cfreq_E}~and~\ref{fig:Cfreq_I} but instead plot the preserved planets if their orbits have been altered by $\Delta e \geq 0.5$ and/or  the semimajor axis changes by more than 50\,percent.

\begin{figure}
    \centering
    \includegraphics[width=\columnwidth]{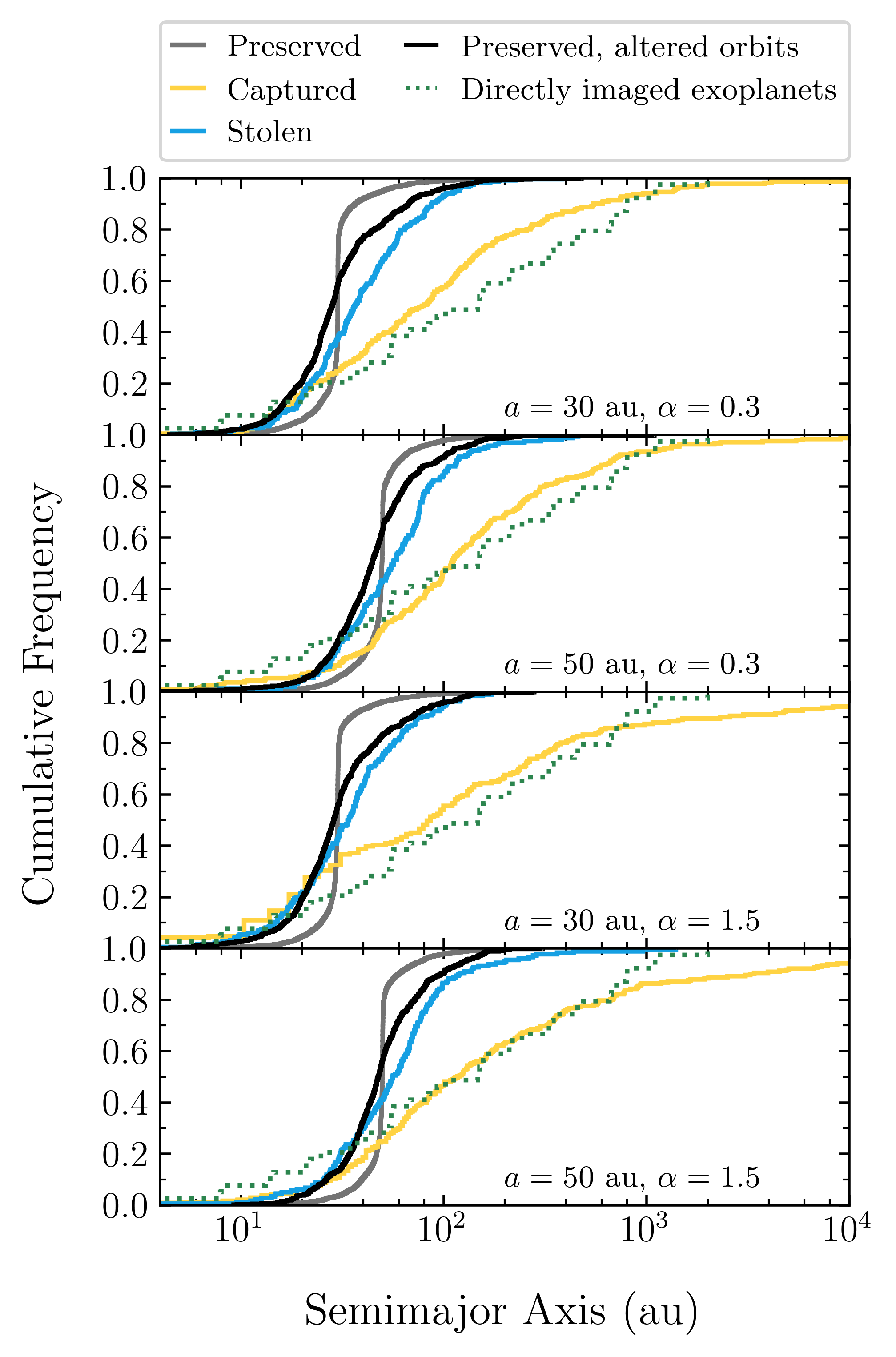}
    \caption{Semimajor axis distributions for planets bound to a star after 10\,Myr, categorised according to the three types of planetary orbit: preserved (grey), captured (yellow), and stolen (blue). We also show the distributions for preserved planets with altered orbits (black), where the orbit is classed as altered if the change in eccentricity is more than $\Delta e = 0.5$ and/or the change in semimajor axis is $\pm$50\,per cent. Results are summed and shown for all 20 realisations of all 4 sets of bound-planet, $r=1$ pc initial conditions, which are shown in separate panels.
The semimajor axis distribution for directly detected exoplanets is also shown for comparison as a dotted green line in each panel.}
    \label{fig:Cfreq_A:app}
\end{figure}

In Fig.~\ref{fig:Cfreq_A:app} the preserved planets with altered orbits (black lines) have a semimajor axis distribution which is much closer to the distribution for the stolen planets (the blue lines) than to the distributions of all preserved planets (the black lines). However, both the stolen and preserved planets can still be distinguished from the captured planets (yellow lines).

\begin{figure}
    \centering
    \includegraphics[width=\columnwidth]{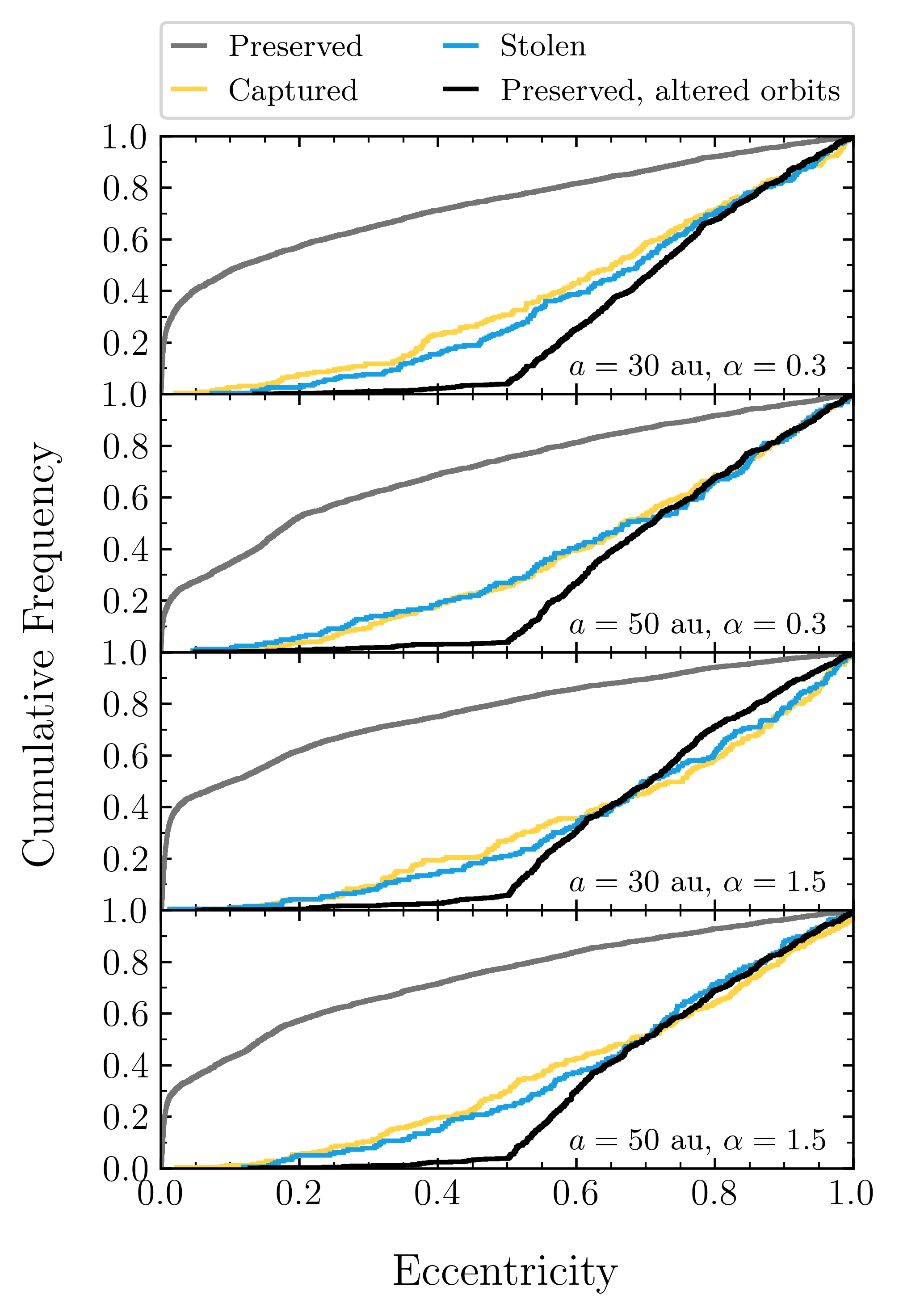}
    \caption{Eccentricity distributions for planets bound to a star after 10\,Myr, categorised according to the three types of planetary orbit: preserved (grey), captured (yellow), and stolen (blue). We also show the distributions for preserved planets with altered orbits (black), where the orbit is classed as altered if the change in eccentricity is more than $\Delta e = 0.5$ and/or the change in semimajor axis is $\pm$50\,per cent. Results are summed and shown for all 20 realisations of all 4 sets of bound-planet, $r=1$ pc initial conditions, which are shown in separate panels.}
    \label{fig:Cfreq_E:app}
\end{figure}

The new constraint that a disrupted planet must have $\Delta e > 0.5$ means that the cumulative distribution of preserved planets with altered orbits is much closer to the eccentricity distributions of the captured and stolen planets (compare the black lines with the blue and yellow lines in Fig.~\ref{fig:Cfreq_E:app}). However, this is somewhat artifical as the captured and stolen planets can have much lower eccentricities than those in the preserved and disrupted distribution.

\begin{figure}
    \centering
    \includegraphics[width=\columnwidth]{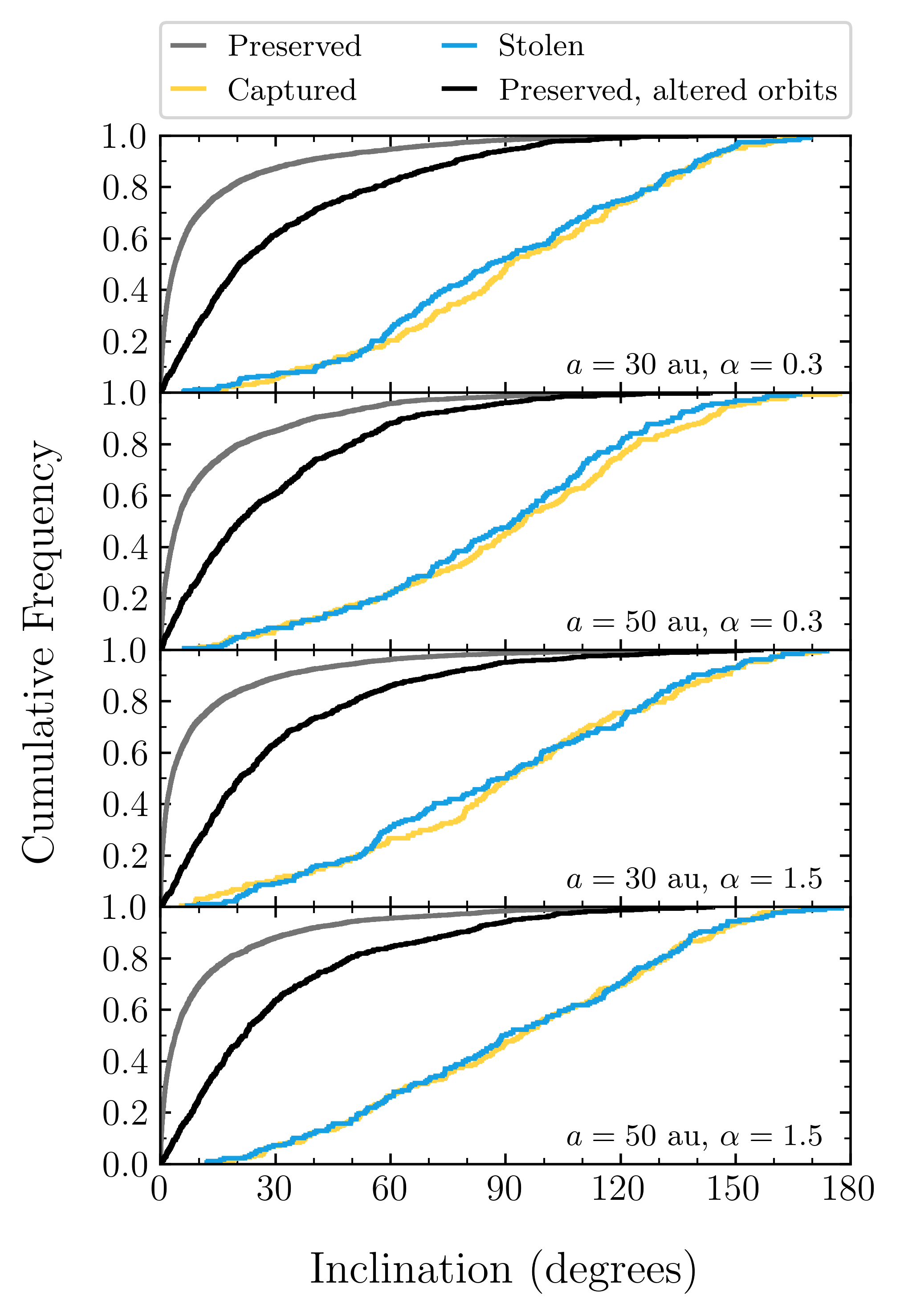}
    \caption{Inclination distributions for planets bound to a star after 10\,Myr, categorised according to the three types of planetary orbit: preserved (grey), captured (yellow), and stolen (blue). We also show the distributions for preserved planets with altered orbits (black), where the orbit is classed as altered if the change in eccentricity is more than $\Delta e = 0.5$ and/or the change in semimajor axis is $\pm$50\,per cent. Results are summed and shown for all 20 realisations of all 4 sets of bound-planet, $r=1$ pc initial conditions, which are shown in separate panels.}
    \label{fig:Cfreq_I:app}
\end{figure}

The (relative) inclination distributions for all of the the preserved planets, and the preserved disrupted planets are most similar, and still easily distinguishable from the inclination distributions of the captured and stolen planets (see Fig.~\ref{fig:Cfreq_I:app}). \\

In summary, the choice of threshold for whether a preserved planet's orbit is disrupted has some bearing on the comparison between preserved planets and captured/stolen planets, although the main results are still discernable.

\bsp	
\label{lastpage}
\end{document}